\documentclass[floatfix, aps, prb, superscriptaddress, reprint]{revtex4-1}

\usepackage{graphicx}               
\usepackage{bm}                     
\usepackage{amsfonts,amssymb}       
\usepackage{dcolumn}                

\newcommand*\pct{\scalebox{.9}{\%}}

\begin{document}

\title{Anisotropic intrinsic lattice thermal conductivity of phosphorene from first principles}

\author{Guangzhao~Qin}
\author{Qing-Bo~Yan}
\email{yan@ucas.ac.cn}
\affiliation{College of Materials Science and Opto-Electronic Technology, University of Chinese Academy of Sciences, Beijing 100049, China}
\author{Zhenzhen~Qin}
\affiliation{College of Electronic Information and Optical Engineering, Nankai University, Tianjin 300071, China}
\author{Sheng-Ying~Yue}
\affiliation{School of Physics, University of Chinese Academy of Sciences, Beijing 100049, China}
\author{Ming~Hu}
\email{hum@ghi.rwth-aachen.de}
\affiliation{Institute of Mineral Engineering, Division of Materials Science and Engineering, Faculty of Georesources and Materials Engineering, RWTH Aachen University, Aachen 52064, Germany}
\affiliation{Aachen Institute for Advanced Study in Computational Engineering Science (AICES), RWTH Aachen University, Aachen 52062, Germany}
\author{Gang~Su}
\email{gsu@ucas.ac.cn}
\homepage{http://tcmp2.ucas.ac.cn/}
\affiliation{School of Physics, University of Chinese Academy of Sciences, Beijing 100049, China}

\date{\today}

\begin{abstract}
Phosphorene, the single layer counterpart of black phosphorus, is a novel
two-dimensional semiconductor with high carrier mobility and a large fundamental
direct band gap, which has attracted tremendous interest recently.
Its potential applications in nano-electronics and thermoelectrics call for a
fundamental study of the phonon transport.
Here, we calculate the intrinsic lattice thermal conductivity of phosphorene by
solving the phonon Boltzmann transport equation (BTE) based on first-principles
calculations.
The thermal conductivity of phosphorene at $300\,\mathrm{K}$ is
$30.15\,\mathrm{Wm^{-1}K^{-1}}$ (zigzag) and $13.65\,\mathrm{Wm^{-1}K^{-1}}$
(armchair), showing an obvious anisotropy along different directions.
The calculated thermal conductivity fits perfectly to the inverse relation with
temperature when the temperature is higher than Debye temperature ($\Theta_D =
278.66\,\mathrm{K}$).
In comparison to graphene, the minor contribution around $5\pct$ of the ZA mode
is responsible for the low thermal conductivity of phosphorene.
In addition, the representative mean free path (MFP), a critical size for phonon
transport, is also obtained.
\end{abstract}

\maketitle

Black phosphorus (BP) has a puckered layered honeycomb structure with layers
held together by \emph{van der Waals} forces, which is similar to graphite.
\cite{PhysRevLett.112.176801, NatCommunBPWeiJi, BP_NATURE}
Few-layer BP has been successfully mechanically exfoliated
\cite{ACSNano84033nn501226z, FIELD_EFFECT_BP} and attracted tremendous interest
recently.
\cite{ACSNano84033nn501226z, FIELD_EFFECT_BP, BPXWJ4458, PhysRevB.90.081408,
C4NR02164A,
apl-104-10-103106,
PhysRevB.89.235319, 
BPAPL1.4885215, BPNCJS4727, 
QINsrep046946, PhysRevB.90.085433, doi:10.1021nl502865s,
PhysRevLett.112.176802,
doi10.1021nl5032293, C4CP03890H}
Phosphorene, the single layer counterpart of BP, is a novel anisotropic
two-dimensional (2D) semiconductor with high carrier mobility
\cite{ACSNano84033nn501226z, FIELD_EFFECT_BP, BPXWJ4458} and a large fundamental
direct band gap $\sim 1.5\,\mathrm{eV}$, \cite{PhysRevB.89.235319} promising
its potential applications in nano-electronics besides the already known 2D
semiconductors such as graphene, germanane, silicene and transition metal
dichalcogenides (TMDCs).\cite{BP_NATURE}
There have already been a lot of theoretical and experimental works exploring
the possible applications of phosphorene as nano-electronic devices, such as
field-effect transistors and photo-transistors.
\cite{C4NR02164A, apl-104-10-103106, FIELD_EFFECT_BP, ACSNano84033nn501226z,
BPXWJ4458, PhysRevB.90.081408}
In addition to extensive studies on its electrical properties, there are also a
lot of theoretical explorations on its potential applications in
thermoelectrics.
\cite{QINsrep046946, doi:10.1021nl502865s, PhysRevB.90.085433}
Phosphorene is found possessing a high \emph{ZT} value, implying it a potential
good thermoelectric material.
\cite{PhysRevB.90.085433, doi:10.1021nl502865s}

All these electrical and thermoelectrical applications of phosphorene are
closely related to its thermal properties.
However, the thermal conductivity of bulk BP was only roughly measured about
fifty years ago\cite{PhysRev.139.A507} and the thermal conductivity of
phosphorene was just simply estimated theoretically
recently\cite{doi:10.1021nl502865s}.
Considering the potential valuable applications of phosphorene as
nano-electronic and thermoelectric devices, it is necessary to fundamentally
study the thermal conductivity and phonon transport in this new 2D
material from first principles.

In this paper, we calculate the intrinsic lattice thermal conductivity of
phosphorene by solving the phonon Boltzmann transport equation (BTE) based on
first-principles calculations.
The thermal conductivity of phosphorene is found anisotropic, and the calculated
thermal conductivity fits perfectly to the inverse relation with temperature
when the temperature is higher than Debye temperature.
Furthermore, we extract the contribution and the relaxation time of each
phonon branch to investigate the underlying mechanism behind the low thermal
conductivity of phosphorene compared to graphene.
At last, we get the representative mean free path (MFP) of phosphorene that is
important for the study of size effect and nano-engineering.

All first-principles calculations are performed based on the density functional
theory (DFT) as implemented in the Vienna \emph{ab-initio} simulation package
(\texttt{\textsc{vasp}})\cite{PhysRevB.54.11169}.
The Perdew-Burke-Ernzerhof (PBE)\cite{PhysRevLett.77.3865} of generalized
gradient approximation (GGA) is chosen as the exchange-correlation functional.
The kinetic energy cutoff of wave functions is set as $700\,\mathrm{eV}$, and a
Monkhorst-Pack \cite{PhysRevB.13.5188} $k$-mesh of $10\times 8\times 1$ is used
to sample the Brillouin Zone (BZ), with the energy convergence threshold set as
$10^{-8}\,\mathrm{eV}$.
A large vacuum spacing of at least $15\,\textrm{\AA}$ is used to hinder
interactions between periodic layers arising from the employed periodic boundary
condition.
All geometries are fully optimized until the maximal Hellmann-Feynman force is
no larger than $10^{-4}\,\textrm{eV/\AA}$.

\begin{figure}[h]
\centering
    \includegraphics[width=0.35\textwidth]{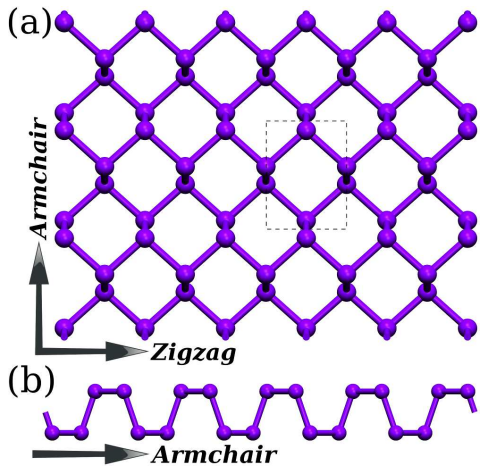}
\caption{\label{fig:structure}
(Color online)
(a) Top view and (b) side view of the monolayer phosphorene.
The unit cell is marked with dashed line.
The zigzag and armchair directions are indicated with arrows.
}
\end{figure}

As shown in Fig.~\ref{fig:structure}, the optimized structure of phosphorene
possesses a hinge-like structure along the armchair direction, which is
distinctly different from the flat graphene and buckled
silicene\cite{QINsrep046946, PhysRevB.82.115427, RevModPhys.81.109,
PhysRevB.89.054310, apl1041310.10631.4870586}.
The obtained lattice constants (zigzag: $3.32\,\textrm{\AA}$, armchair:
$4.58\,\textrm{\AA}$) are in good agreement with previous
studies\cite{NatCommunBPWeiJi, doi:10.1021/nl500935z}.
In one unit cell, there are two atoms per layer and four atoms in total, which
means 3 acoustic and 9 optical phonon branches.

\begin{figure}[h]
\centering
    \includegraphics[width=0.45\textwidth]{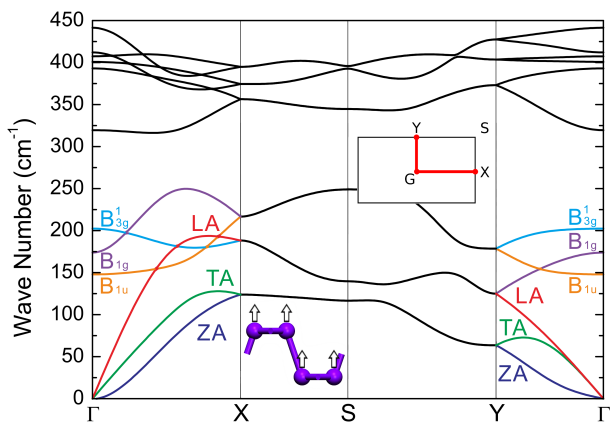}
\caption{\label{fig:dispersion}
(Color online)
Phonon dispersion along the path passing through the main high-symmetry $k$-points in BZ of phosphorene.
The acoustic phonon branches (ZA, TA and LA) and three optical phonon branches
($\mathrm{B_{1u}}$, $\mathrm{B_{1g}}$ and $\mathrm{B_{3g}^1}$) are indicated
with different colors.
The BZ with high symmetry $k$-points indicated and the vibrational direction of
the ZA mode are shown in the insets.
}
\end{figure}

In the calculation of phonon dispersion, the $5\times 5\times 1$ supercell
containing 100 atoms is constructed to ensure the convergence.
The Monkhorst-Pack $k$-mesh of $2\times 2\times 1$ is used here to sample the
BZ.
The harmonic second order interatomic force constants (IFCs) are obtained within
the linear response framework by employing the density functional perturbation
theory (DFPT) as implemented in the \texttt{\textsc{vasp}}
code\cite{PhysRevB.54.11169}.
Then we could get the phonon dispersion of phosphorene using the
\texttt{\textsc{phonopy}} package\cite{phonopy} based on the harmonic second
order IFCs.
The obtained phonon dispersion, as shown in Fig.~\ref{fig:dispersion}, is in
consistent with other works\cite{PhysRevLett.112.176802, doi:10.1021nl502865s}.

Similar to graphene and silicene \cite{PhysRevB.82.115427, RevModPhys.81.109,
PhysRevB.89.054310, apl1041310.10631.4870586}, phosphorene also has a quadratic
flexural phonon branch ($z$-direction acoustic mode, ZA) near the $\Gamma$
point, which is a typical feature of 2D materials\cite{RevModPhys.81.109}.
The vibrational direction of the ZA mode is exactly perpendicular to the plane,
i.e.\ along the $z$ direction, which is similar to graphene but different from
silicene whose flexural phonon mode is not purely out-of-plane
vibration\cite{PhysRevB.89.054310, apl1041310.10631.4870586}.
Based on the slopes of the longitudinal acoustic (LA) branch near $\Gamma$
point, we could get the group velocities along $\Gamma$-X (zigzag) and
$\Gamma$-Y (armchair) as $7.83\,\mathrm{km/s}$ and $4.01\,\mathrm{km/s}$,
respectively, which are in good agreement with previous
results\cite{PhysRevLett.112.176802}.
With the highest frequency of normal mode vibration (Debye frequency) $\nu_m =
5.81\,\mathrm{THz}$, the Debye temperature ($\Theta_D$) could be calculated:
$\Theta_D = h\nu_m/k_B = 278.66\,\mathrm{K}$, where $h$ is the Planck's constant
and $k_B$ is the Boltzmann constant.

\begin{figure}[h]
\centering
    \includegraphics[width=0.45\textwidth]{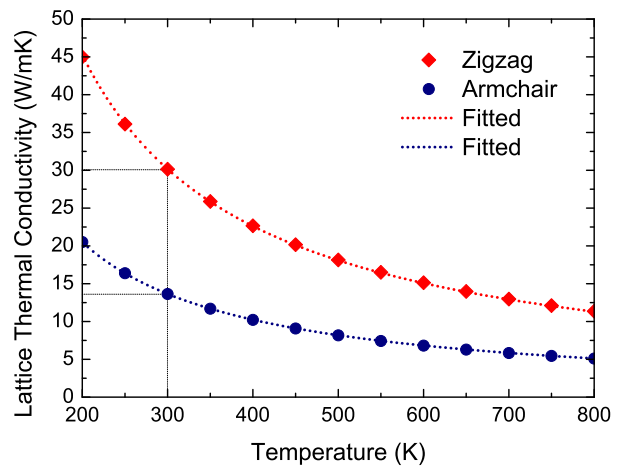}
\caption{\label{fig:conductivity}
(Color online)
Calculated lattice thermal conductivity ($\kappa$) of phosphorene along zigzag
(red diamond) and armchair (blue circle) directions as a function of temperature
ranging from $200\,\mathrm{K}$ to $800\,\mathrm{K}$.
The thermal conductivity fitted by the inverse relation with temperature
($\kappa \sim 1/T$) is plotted with dot lines.
}
\end{figure}

For the calculation of lattice thermal conductivity, anharmonic third order IFCs
are also necessary besides the harmonic second order IFCs obtained above.
The same $5\times 5\times 1$ supercell and $2\times 2\times 1$ Monkhorst-Pack
$k$-mesh are used to get the anharmonic third order IFCs, and interactions are
taken into account up to the fourth nearest neighbors\cite{ap1.2822891,
PhysRevB.86.174307}.
We additionally obtain the dielectric tensor and Born effective charges
based on DFPT for taking into account long-range electrostatic interactions.
As a thickness is necessary for the calculation of thermal conductivity for
2D materials, we choose the half of the length of bulk lattice constant along
$z$ direction as the thickness of phosphorene, which is $5.36\,\textrm{\AA}$.
\cite{QINsrep046946, PhysRevB.89.155426}
Only considering the phonon-phonon scattering processes, the intrinsic lattice
thermal conductivity could be obtained by solving the phonon BTE 
as implemented in the \texttt{\textsc{ShengBTE}} code\cite{Li20141747}.
It yields predictive parameter free estimate of thermal conductivity using only
basic information of the chemical structure.
Detailed information on the workflow could be found in the Supplemental
Material.

The intrinsic lattice thermal conductivity ($\kappa$) of phosphorene, as shown
in Fig.~\ref{fig:conductivity}, is obviously anisotropic that the thermal
conductivity along the zigzag direction is generally twice the thermal
conductivity along the armchair direction, which coincides with the previous
expectation\cite{doi:10.1021nl502865s}.
The anisotropy lies in the asymmetry of group velocities along the $\Gamma$-X
(zigzag) and $\Gamma$-Y (armchair) directions, which may be due to the
anisotropic hinge-like structure of phosphorene\cite{QINsrep046946,
BPAPL1.4885215}.
Similar anisotropy of thermal conductivity is also found in SnSe, which
possesses almost the same hinge-like structure as
phosphorene.\cite{apl.10510101907.1.4895770}

The thermal conductivity of phosphorene at $300\,\mathrm{K}$ is
$30.15\,\mathrm{Wm^{-1}K^{-1}}$ (zigzag) and $13.65\,\mathrm{Wm^{-1}K^{-1}}$
(armchair).
Note that the calculated thermal conductivity along the armchair direction is
close to the reported experimental value ($12.1\,\mathrm{Wm^{-1}K^{-1}}$, with
no direction information) of bulk BP \cite{PhysRev.139.A507} while being a
little larger, which may be due to the effect of interlayer interactions.
The effect of long-range electrostatic interactions introduced from dielectric
tensor and Born effective charges is also examined.
It is found that the lack of long-range electrostatic interactions would only
make the calculated thermal conductivity around $4\pct$ smaller.
Furthermore, we fit the calculated thermal conductivity along zigzag and
armchair directions, respectively, and find that they both satisfy an inverse
relation with temperature, i.e.\ $\kappa \sim 1/T$.
The fitted thermal conductivity is also plotted with dot lines in
Fig.~\ref{fig:conductivity}.
It is obvious that the fitted thermal conductivity coincides perfectly with the
calculated thermal conductivity.
The inverse relation of thermal conductivity with temperature that also lies in
other materials such as PbSe and Mg$_2$(Si,Sn) \cite{PhysRevB.82.035204,
PhysRevB.86.155204} has been often observed when the temperature is higher than
Debye temperature ($\Theta_D$), which could be explained by the phonon
scattering mechanism in pure semiconductors\cite{bookTCPOSALT}.

\begin{figure}[h]
\centering
    \includegraphics[width=0.45\textwidth]{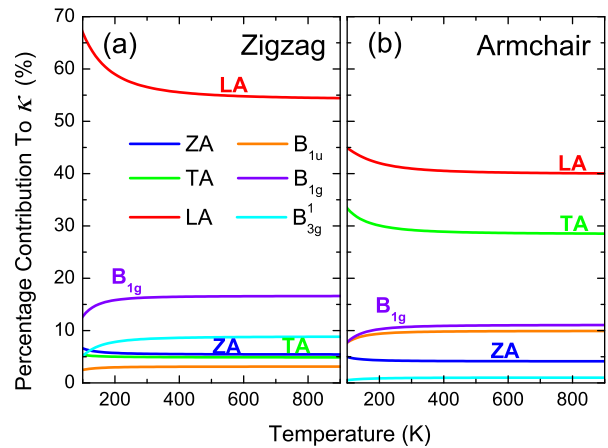}
\caption{\label{fig:contribution}
(Color online)
The percentage contribution of each phonon branch to thermal conductivity along
(a) zigzag and (b) armchair directions as a function of temperature.
Those optical phonon branches contributing less than $1\protect\pct$ are
not shown.
}
\end{figure}

The thermal conductivity of phosphorene is on the same order of magnitude as
that of silicene ($5-65\,\mathrm{Wm^{-1}K^{-1}}$)\cite{PhysRevB.89.054310,
PhysRevB.87.195417, apl1041310.10631.4870586, arXiv:1404.2874}, while at least
two orders of magnitude lower than that of graphene
($3000-5000\,\mathrm{Wm^{-1}K^{-1}}$)\cite{PhysRevB.82.115427,
PhysRevB.89.155426}.
In order to understand the underlying mechanism, we examine the contributions of
different phonon branches to the thermal conductivity of phosphorene in both
zigzag and armchair directions, as shown in Fig.~\ref{fig:contribution}.
The contribution of the ZA mode is around $5\pct$ at room temperature, which is
close to that of silicene ($7.5\pct$) but much smaller than that of graphene
($75\pct$)\cite{arXiv:1404.2874, PhysRevB.89.155426}.
Hence, the reason for the low thermal conductivity of phosphorene might be the
same as silicene that is assumed to be due to a small contribution of the ZA
mode.
Comparing with graphene and silicene, the symmetry-based phonon-phonon
scattering selection rule \cite{PhysRevB.82.115427} is broke by the puckered
hinge-like structure of phosphorene, resulting in a large scattering rate of the
out-of-plane ZA mode, which thus leads to its small contribution to the thermal
conductivity.
Note that when temperature increases, the contributions of optical phonon
branches increase while the contributions of acoustic phonon branches decrease,
which might be due to the fact that a high temperature can excite high frequency
optical phonon modes that represent the collective opposite vibrations of atoms.

\begin{figure}[h]
\centering
    \includegraphics[width=0.45\textwidth]{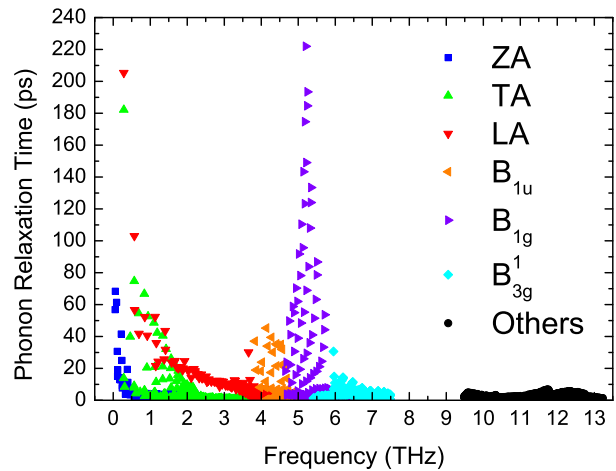}
\caption{\label{fig:relaxationtime}
(Color online)
Phonon relaxation time of each phonon branch as a function of frequency.
}
\end{figure}

To gain more information on the contributions of different phonon branches to
the thermal conductivity, we extract the relaxation time of each phonon branch
as a function of frequency, as shown in Fig.~\ref{fig:relaxationtime}.
The relaxation time of ZA mode is shorter than that of LA mode and TA mode,
which is due to the large scattering rate of ZA mode as discussed above.
Thus the ZA mode contributes little to the thermal conductivity, leading to the
low thermal conductivity of phosphorene compared to graphene.
It can also be seen that the relaxation time of LA mode is the longest among the
three acoustic phonon branches, especially for phonon frequency between $2$ and
$4\,\mathrm{THz}$. 
Together with its highest group velocity, it is understandable that the LA mode
contributes the most to the thermal conductivity as shown in
Fig.~\ref{fig:contribution}.
Note that the contribution of TA mode is rather large along the armchair
direction as shown in Fig.~\ref{fig:contribution}(b), which is due to almost the
same group velocity of TA mode as LA mode along the $\Gamma$-Y (armchair)
direction as shown in Fig.~\ref{fig:dispersion}.
Considering the very long relaxation time of the optical phonon branch
$\mathrm{B_{1g}}$, the $\mathrm{B_{1g}}$ mode contributes much more than other
optical phonon branches to the thermal conductivity.

\begin{figure}[h]
\centering
    \includegraphics[width=0.45\textwidth]{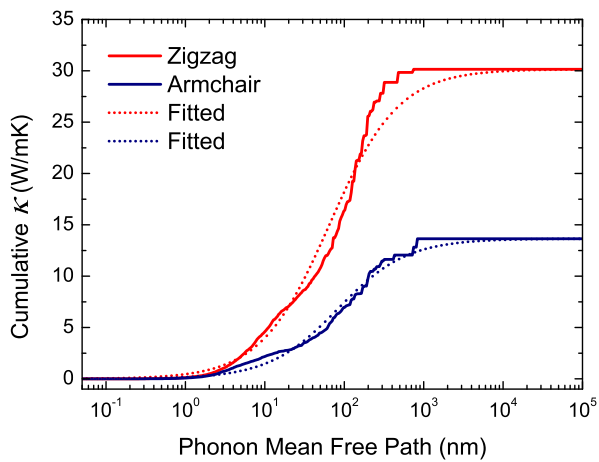}
\caption{\label{fig:cumulative}
(Color online)
Cumulative lattice thermal conductivity of phosphorene along zigzag (red) and
armchair (blue) directions as a function of the phonon MFP at $300\,\mathrm{K}$.
The curves fitted by the function presented in text are plotted with dot lines.
}
\end{figure}

The cumulative thermal conductivity ($\kappa$) with respect to the phonon MFP at
$300\,\mathrm{K}$ for both zigzag and armchair directions are plotted in
Fig.~\ref{fig:cumulative}.
We fit the data to a single parametric function \cite{Li20141747}
$$\kappa(l\leq l_{max}) = \frac{\kappa_0}{1+l_0/l_{max}}\ ,$$
where $\kappa_0$ and $l_{max}$ are the ultimate cumulated thermal conductivity
and the maximal MFP concerned, respectively, and $l_0$ is the parameter to be
evaluated.
The fitted curves, as plotted with dot lines in Fig.~\ref{fig:cumulative},
reproduce the calculated data quite well and yield the parameter $l_0$ as
$66\,\mathrm{nm}$ and $83\,\mathrm{nm}$ for zigzag and armchair directions,
respectively, which could be interpreted as the representative MFP of
phosphorene.
The representative MFP is helpful for the study of the size effect on the
ballistic or diffusive phonon transport.
A related quantity is the ratio of thermal conductivity to the thermal
conductivity per unit of MFP in the small grain limit, which is an estimate of
the characteristic size below which the nanostructuring induced phonon scattering
dominates over the anharmonic phonon-phonon scattering.
This is critical to thermal design with nanostructuring in that the thermal
conductivity could be modulated effectively by nanostructuring when the
nanostructure size below the characteristic size.
The obtained values for phosphorene are $17\,\mathrm{nm}$ (zigzag) and
$15\,\mathrm{nm}$ (armchair).

In summary, we have calculated the intrinsic lattice thermal conductivity of
phosphorene from first principles.
The thermal conductivity of phosphorene at $300\,\mathrm{K}$ is found to be
$30.15\,\mathrm{Wm^{-1}K^{-1}}$ (zigzag) and $13.65\,\mathrm{Wm^{-1}K^{-1}}$
(armchair), which is anisotropic along different directions.
The calculated thermal conductivity fits perfectly to the inverse relation with
temperature when the temperature is higher than Debye temperature ($\Theta_D =
278.66\,\mathrm{K}$).
Comparing with graphene, the minor contribution around $5\pct$ of the ZA mode is
responsible for the low thermal conductivity of phosphorene, which might be due
to the fact that the symmetry-based phonon-phonon scattering selection rule is
broke by the puckered hinge-like structure of phosphorene.
In addition, the representative MFP, a critical size for phonon transport, is
also obtained along zigzag and armchair directions.

The authors would like to thank Yingying Li at ZJU, Chao-Sheng Lian at IOP of
CAS, Hui-Juan Cui and Prof.\ Qing-Rong Zheng at UCAS for helpful discussions.
This work is supported in part by the NSFC (Grant No.\ 11004239), the MOST
(Grant No.\ 2012CB932901 and No.\ 2013CB933401) of China, and the fund from CAS.
All calculations are performed on Nebulae (DAWN6000) in National Supercomputing
Center in Shenzhen and MagicCube (DAWN5000A) in Shanghai Supercomputer Center,
China.

\bibliography{bibliography.bib}

\begin{thebibliography}{38}%
\makeatletter
\providecommand \@ifxundefined [1]{%
 \@ifx{#1\undefined}
}%
\providecommand \@ifnum [1]{%
 \ifnum #1\expandafter \@firstoftwo
 \else \expandafter \@secondoftwo
 \fi
}%
\providecommand \@ifx [1]{%
 \ifx #1\expandafter \@firstoftwo
 \else \expandafter \@secondoftwo
 \fi
}%
\providecommand \natexlab [1]{#1}%
\providecommand \enquote  [1]{``#1''}%
\providecommand \bibnamefont  [1]{#1}%
\providecommand \bibfnamefont [1]{#1}%
\providecommand \citenamefont [1]{#1}%
\providecommand \href@noop [0]{\@secondoftwo}%
\providecommand \href [0]{\begingroup \@sanitize@url \@href}%
\providecommand \@href[1]{\@@startlink{#1}\@@href}%
\providecommand \@@href[1]{\endgroup#1\@@endlink}%
\providecommand \@sanitize@url [0]{\catcode `\\12\catcode `\$12\catcode
  `\&12\catcode `\#12\catcode `\^12\catcode `\_12\catcode `\%12\relax}%
\providecommand \@@startlink[1]{}%
\providecommand \@@endlink[0]{}%
\providecommand \url  [0]{\begingroup\@sanitize@url \@url }%
\providecommand \@url [1]{\endgroup\@href {#1}{\urlprefix }}%
\providecommand \urlprefix  [0]{URL }%
\providecommand \Eprint [0]{\href }%
\providecommand \doibase [0]{http://dx.doi.org/}%
\providecommand \selectlanguage [0]{\@gobble}%
\providecommand \bibinfo  [0]{\@secondoftwo}%
\providecommand \bibfield  [0]{\@secondoftwo}%
\providecommand \translation [1]{[#1]}%
\providecommand \BibitemOpen [0]{}%
\providecommand \bibitemStop [0]{}%
\providecommand \bibitemNoStop [0]{.\EOS\space}%
\providecommand \EOS [0]{\spacefactor3000\relax}%
\providecommand \BibitemShut  [1]{\csname bibitem#1\endcsname}%
\let\auto@bib@innerbib\@empty
\bibitem [{\citenamefont {Rodin}\ \emph {et~al.}(2014)\citenamefont {Rodin},
  \citenamefont {Carvalho},\ and\ \citenamefont
  {Castro~Neto}}]{PhysRevLett.112.176801}%
  \BibitemOpen
  \bibfield  {author} {\bibinfo {author} {\bibfnamefont {A.~S.}\ \bibnamefont
  {Rodin}}, \bibinfo {author} {\bibfnamefont {A.}~\bibnamefont {Carvalho}}, \
  and\ \bibinfo {author} {\bibfnamefont {H.}~\bibnamefont {Castro~Neto},
  \bibfnamefont {A.}},\ }\href {\doibase 10.1103/PhysRevLett.112.176801}
  {\bibfield  {journal} {\bibinfo  {journal} {Phys. Rev. Lett.}\ }\textbf
  {\bibinfo {volume} {112}},\ \bibinfo {pages} {176801} (\bibinfo {year}
  {2014})}\BibitemShut {NoStop}%
\bibitem [{\citenamefont {Qiao}\ \emph {et~al.}(2014)\citenamefont {Qiao},
  \citenamefont {Kong}, \citenamefont {Hu}, \citenamefont {Yang},\ and\
  \citenamefont {Ji}}]{NatCommunBPWeiJi}%
  \BibitemOpen
  \bibfield  {author} {\bibinfo {author} {\bibfnamefont {J.}~\bibnamefont
  {Qiao}}, \bibinfo {author} {\bibfnamefont {X.}~\bibnamefont {Kong}}, \bibinfo
  {author} {\bibfnamefont {Z.-X.}\ \bibnamefont {Hu}}, \bibinfo {author}
  {\bibfnamefont {F.}~\bibnamefont {Yang}}, \ and\ \bibinfo {author}
  {\bibfnamefont {W.}~\bibnamefont {Ji}},\ }\href
  {http://www.nature.com/ncomms/2014/140721/ncomms5475/full/ncomms5475.html}
  {\bibfield  {journal} {\bibinfo  {journal} {Nat. Commun.}\ }\textbf {\bibinfo
  {volume} {5}},\ \bibinfo {pages} {4475} (\bibinfo {year} {2014})}\BibitemShut
  {NoStop}%
\bibitem [{\citenamefont {Churchill}\ and\ \citenamefont
  {Jarillo-Herrero}(2014)}]{BP_NATURE}%
  \BibitemOpen
  \bibfield  {author} {\bibinfo {author} {\bibfnamefont {H.~O.~H.}\
  \bibnamefont {Churchill}}\ and\ \bibinfo {author} {\bibfnamefont
  {P.}~\bibnamefont {Jarillo-Herrero}},\ }\href
  {http://dx.doi.org/10.1038/nnano.2014.85} {\bibfield  {journal} {\bibinfo
  {journal} {Nature Nanotech.}\ }\textbf {\bibinfo {volume} {9}},\ \bibinfo
  {pages} {330} (\bibinfo {year} {2014})}\BibitemShut {NoStop}%
\bibitem [{\citenamefont {Liu}\ \emph {et~al.}(2014)\citenamefont {Liu},
  \citenamefont {Neal}, \citenamefont {Zhu}, \citenamefont {Luo}, \citenamefont
  {Xu}, \citenamefont {Tománek},\ and\ \citenamefont
  {Ye}}]{ACSNano84033nn501226z}%
  \BibitemOpen
  \bibfield  {author} {\bibinfo {author} {\bibfnamefont {H.}~\bibnamefont
  {Liu}}, \bibinfo {author} {\bibfnamefont {A.~T.}\ \bibnamefont {Neal}},
  \bibinfo {author} {\bibfnamefont {Z.}~\bibnamefont {Zhu}}, \bibinfo {author}
  {\bibfnamefont {Z.}~\bibnamefont {Luo}}, \bibinfo {author} {\bibfnamefont
  {X.}~\bibnamefont {Xu}}, \bibinfo {author} {\bibfnamefont {D.}~\bibnamefont
  {Tománek}}, \ and\ \bibinfo {author} {\bibfnamefont {P.~D.}\ \bibnamefont
  {Ye}},\ }\href {\doibase 10.1021/nn501226z} {\bibfield  {journal} {\bibinfo
  {journal} {ACS Nano}\ }\textbf {\bibinfo {volume} {8}},\ \bibinfo {pages}
  {4033} (\bibinfo {year} {2014})}\BibitemShut {NoStop}%
\bibitem [{\citenamefont {Li}\ \emph {et~al.}(2014{\natexlab{a}})\citenamefont
  {Li}, \citenamefont {Yu}, \citenamefont {Ye}, \citenamefont {Ge},
  \citenamefont {Ou}, \citenamefont {Wu}, \citenamefont {Feng}, \citenamefont
  {Chen},\ and\ \citenamefont {Zhang}}]{FIELD_EFFECT_BP}%
  \BibitemOpen
  \bibfield  {author} {\bibinfo {author} {\bibfnamefont {L.}~\bibnamefont
  {Li}}, \bibinfo {author} {\bibfnamefont {Y.}~\bibnamefont {Yu}}, \bibinfo
  {author} {\bibfnamefont {G.~J.}\ \bibnamefont {Ye}}, \bibinfo {author}
  {\bibfnamefont {Q.}~\bibnamefont {Ge}}, \bibinfo {author} {\bibfnamefont
  {X.}~\bibnamefont {Ou}}, \bibinfo {author} {\bibfnamefont {H.}~\bibnamefont
  {Wu}}, \bibinfo {author} {\bibfnamefont {D.}~\bibnamefont {Feng}}, \bibinfo
  {author} {\bibfnamefont {X.~H.}\ \bibnamefont {Chen}}, \ and\ \bibinfo
  {author} {\bibfnamefont {Y.}~\bibnamefont {Zhang}},\ }\href
  {http://dx.doi.org/10.1038/nnano.2014.35} {\bibfield  {journal} {\bibinfo
  {journal} {Nature Nanotech.}\ }\textbf {\bibinfo {volume} {9}},\ \bibinfo
  {pages} {372} (\bibinfo {year} {2014}{\natexlab{a}})}\BibitemShut {NoStop}%
\bibitem [{\citenamefont {Xia}\ \emph {et~al.}(2014)\citenamefont {Xia},
  \citenamefont {Wang},\ and\ \citenamefont {Jia}}]{BPXWJ4458}%
  \BibitemOpen
  \bibfield  {author} {\bibinfo {author} {\bibfnamefont {F.}~\bibnamefont
  {Xia}}, \bibinfo {author} {\bibfnamefont {H.}~\bibnamefont {Wang}}, \ and\
  \bibinfo {author} {\bibfnamefont {Y.}~\bibnamefont {Jia}},\ }\href
  {http://www.nature.com/ncomms/2014/140721/ncomms5458/abs/ncomms5458.html}
  {\bibfield  {journal} {\bibinfo  {journal} {Nat. Commun.}\ }\textbf {\bibinfo
  {volume} {5}},\ \bibinfo {pages} {4458} (\bibinfo {year} {2014})}\BibitemShut
  {NoStop}%
\bibitem [{\citenamefont {Low}\ \emph {et~al.}(2014)\citenamefont {Low},
  \citenamefont {Engel}, \citenamefont {Steiner},\ and\ \citenamefont
  {Avouris}}]{PhysRevB.90.081408}%
  \BibitemOpen
  \bibfield  {author} {\bibinfo {author} {\bibfnamefont {T.}~\bibnamefont
  {Low}}, \bibinfo {author} {\bibfnamefont {M.}~\bibnamefont {Engel}}, \bibinfo
  {author} {\bibfnamefont {M.}~\bibnamefont {Steiner}}, \ and\ \bibinfo
  {author} {\bibfnamefont {P.}~\bibnamefont {Avouris}},\ }\href {\doibase
  10.1103/PhysRevB.90.081408} {\bibfield  {journal} {\bibinfo  {journal} {Phys.
  Rev. B}\ }\textbf {\bibinfo {volume} {90}},\ \bibinfo {pages} {081408}
  (\bibinfo {year} {2014})}\BibitemShut {NoStop}%
\bibitem [{\citenamefont {Hong}\ \emph {et~al.}(2014)\citenamefont {Hong},
  \citenamefont {Chamlagain}, \citenamefont {Lin}, \citenamefont {Chuang},
  \citenamefont {Pan}, \citenamefont {Zhou},\ and\ \citenamefont
  {Xu}}]{C4NR02164A}%
  \BibitemOpen
  \bibfield  {author} {\bibinfo {author} {\bibfnamefont {T.}~\bibnamefont
  {Hong}}, \bibinfo {author} {\bibfnamefont {B.}~\bibnamefont {Chamlagain}},
  \bibinfo {author} {\bibfnamefont {W.}~\bibnamefont {Lin}}, \bibinfo {author}
  {\bibfnamefont {H.-J.}\ \bibnamefont {Chuang}}, \bibinfo {author}
  {\bibfnamefont {M.}~\bibnamefont {Pan}}, \bibinfo {author} {\bibfnamefont
  {Z.}~\bibnamefont {Zhou}}, \ and\ \bibinfo {author} {\bibfnamefont {Y.-Q.}\
  \bibnamefont {Xu}},\ }\href {\doibase 10.1039/C4NR02164A} {\bibfield
  {journal} {\bibinfo  {journal} {Nanoscale}\ }\textbf {\bibinfo {volume}
  {6}},\ \bibinfo {pages} {8978} (\bibinfo {year} {2014})}\BibitemShut
  {NoStop}%
\bibitem [{\citenamefont {Koenig}\ \emph {et~al.}(2014)\citenamefont {Koenig},
  \citenamefont {Doganov}, \citenamefont {Schmidt}, \citenamefont
  {Castro~Neto},\ and\ \citenamefont {Özyilmaz}}]{apl-104-10-103106}%
  \BibitemOpen
  \bibfield  {author} {\bibinfo {author} {\bibfnamefont {S.~P.}\ \bibnamefont
  {Koenig}}, \bibinfo {author} {\bibfnamefont {R.~A.}\ \bibnamefont {Doganov}},
  \bibinfo {author} {\bibfnamefont {H.}~\bibnamefont {Schmidt}}, \bibinfo
  {author} {\bibfnamefont {A.~H.}\ \bibnamefont {Castro~Neto}}, \ and\ \bibinfo
  {author} {\bibfnamefont {B.}~\bibnamefont {Özyilmaz}},\ }\href {\doibase
  http://dx.doi.org/10.1063/1.4868132} {\bibfield  {journal} {\bibinfo
  {journal} {Appl. Phys. Lett.}\ }\textbf {\bibinfo {volume} {104}},\ \bibinfo
  {eid} {103106} (\bibinfo {year} {2014})}\BibitemShut {NoStop}%
\bibitem [{\citenamefont {Tran}\ \emph {et~al.}(2014)\citenamefont {Tran},
  \citenamefont {Soklaski}, \citenamefont {Liang},\ and\ \citenamefont
  {Yang}}]{PhysRevB.89.235319}%
  \BibitemOpen
  \bibfield  {author} {\bibinfo {author} {\bibfnamefont {V.}~\bibnamefont
  {Tran}}, \bibinfo {author} {\bibfnamefont {R.}~\bibnamefont {Soklaski}},
  \bibinfo {author} {\bibfnamefont {Y.}~\bibnamefont {Liang}}, \ and\ \bibinfo
  {author} {\bibfnamefont {L.}~\bibnamefont {Yang}},\ }\href {\doibase
  10.1103/PhysRevB.89.235319} {\bibfield  {journal} {\bibinfo  {journal} {Phys.
  Rev. B}\ }\textbf {\bibinfo {volume} {89}},\ \bibinfo {pages} {235319}
  (\bibinfo {year} {2014})}\BibitemShut {NoStop}%
\bibitem [{\citenamefont {Wei}\ and\ \citenamefont
  {Peng}(2014)}]{BPAPL1.4885215}%
  \BibitemOpen
  \bibfield  {author} {\bibinfo {author} {\bibfnamefont {Q.}~\bibnamefont
  {Wei}}\ and\ \bibinfo {author} {\bibfnamefont {X.}~\bibnamefont {Peng}},\
  }\href {\doibase http://dx.doi.org/10.1063/1.4885215} {\bibfield  {journal}
  {\bibinfo  {journal} {Appl. Phys. Lett.}\ }\textbf {\bibinfo {volume}
  {104}},\ \bibinfo {eid} {251915} (\bibinfo {year} {2014})}\BibitemShut
  {NoStop}%
\bibitem [{\citenamefont {Jiang}\ and\ \citenamefont
  {Park}(2014)}]{BPNCJS4727}%
  \BibitemOpen
  \bibfield  {author} {\bibinfo {author} {\bibfnamefont {J.-W.}\ \bibnamefont
  {Jiang}}\ and\ \bibinfo {author} {\bibfnamefont {H.~S.}\ \bibnamefont
  {Park}},\ }\href@noop {} {\bibfield  {journal} {\bibinfo  {journal} {Nat.
  Commun.}\ }\textbf {\bibinfo {volume} {5}},\ \bibinfo {pages} {4727}
  (\bibinfo {year} {2014})}\BibitemShut {NoStop}%
\bibitem [{\citenamefont {Qin}\ \emph {et~al.}(2014)\citenamefont {Qin},
  \citenamefont {Yan}, \citenamefont {Qin}, \citenamefont {Yue}, \citenamefont
  {Cui}, \citenamefont {Zheng},\ and\ \citenamefont {Su}}]{QINsrep046946}%
  \BibitemOpen
  \bibfield  {author} {\bibinfo {author} {\bibfnamefont {G.}~\bibnamefont
  {Qin}}, \bibinfo {author} {\bibfnamefont {Q.-B.}\ \bibnamefont {Yan}},
  \bibinfo {author} {\bibfnamefont {Z.}~\bibnamefont {Qin}}, \bibinfo {author}
  {\bibfnamefont {S.-Y.}\ \bibnamefont {Yue}}, \bibinfo {author} {\bibfnamefont
  {H.-J.}\ \bibnamefont {Cui}}, \bibinfo {author} {\bibfnamefont {Q.-R.}\
  \bibnamefont {Zheng}}, \ and\ \bibinfo {author} {\bibfnamefont
  {G.}~\bibnamefont {Su}},\ }\href@noop {} {\bibfield  {journal} {\bibinfo
  {journal} {Scientific Reports}\ }\textbf {\bibinfo {volume} {4}},\ \bibinfo
  {eid} {6946} (\bibinfo {year} {2014})}\BibitemShut {NoStop}%
\bibitem [{\citenamefont {Lv}\ \emph {et~al.}(2014)\citenamefont {Lv},
  \citenamefont {Lu}, \citenamefont {Shao},\ and\ \citenamefont
  {Sun}}]{PhysRevB.90.085433}%
  \BibitemOpen
  \bibfield  {author} {\bibinfo {author} {\bibfnamefont {H.~Y.}\ \bibnamefont
  {Lv}}, \bibinfo {author} {\bibfnamefont {W.~J.}\ \bibnamefont {Lu}}, \bibinfo
  {author} {\bibfnamefont {D.~F.}\ \bibnamefont {Shao}}, \ and\ \bibinfo
  {author} {\bibfnamefont {Y.~P.}\ \bibnamefont {Sun}},\ }\href {\doibase
  10.1103/PhysRevB.90.085433} {\bibfield  {journal} {\bibinfo  {journal} {Phys.
  Rev. B}\ }\textbf {\bibinfo {volume} {90}},\ \bibinfo {pages} {085433}
  (\bibinfo {year} {2014})}\BibitemShut {NoStop}%
\bibitem [{\citenamefont {Fei}\ \emph {et~al.}(2014)\citenamefont {Fei},
  \citenamefont {Faghaninia}, \citenamefont {Soklaski}, \citenamefont {Yan},
  \citenamefont {Lo},\ and\ \citenamefont {Yang}}]{doi:10.1021nl502865s}%
  \BibitemOpen
  \bibfield  {author} {\bibinfo {author} {\bibfnamefont {R.}~\bibnamefont
  {Fei}}, \bibinfo {author} {\bibfnamefont {A.}~\bibnamefont {Faghaninia}},
  \bibinfo {author} {\bibfnamefont {R.}~\bibnamefont {Soklaski}}, \bibinfo
  {author} {\bibfnamefont {J.-A.}\ \bibnamefont {Yan}}, \bibinfo {author}
  {\bibfnamefont {C.}~\bibnamefont {Lo}}, \ and\ \bibinfo {author}
  {\bibfnamefont {L.}~\bibnamefont {Yang}},\ }\href {\doibase
  10.1021/nl502865s} {\bibfield  {journal} {\bibinfo  {journal} {Nano Letters}\
  }\textbf {\bibinfo {volume} {14}},\ \bibinfo {pages} {6393} (\bibinfo {year}
  {2014})},\ \bibinfo {note} {pMID: 25254626},\ \Eprint
  {http://arxiv.org/abs/http://dx.doi.org/10.1021/nl502865s}
  {http://dx.doi.org/10.1021/nl502865s} \BibitemShut {NoStop}%
\bibitem [{\citenamefont {Zhu}\ and\ \citenamefont
  {Tom\'anek}(2014)}]{PhysRevLett.112.176802}%
  \BibitemOpen
  \bibfield  {author} {\bibinfo {author} {\bibfnamefont {Z.}~\bibnamefont
  {Zhu}}\ and\ \bibinfo {author} {\bibfnamefont {D.}~\bibnamefont
  {Tom\'anek}},\ }\href {\doibase 10.1103/PhysRevLett.112.176802} {\bibfield
  {journal} {\bibinfo  {journal} {Phys. Rev. Lett.}\ }\textbf {\bibinfo
  {volume} {112}},\ \bibinfo {pages} {176802} (\bibinfo {year}
  {2014})}\BibitemShut {NoStop}%
\bibitem [{\citenamefont {Wood}\ \emph {et~al.}(2014)\citenamefont {Wood},
  \citenamefont {Wells}, \citenamefont {Jariwala}, \citenamefont {Chen},
  \citenamefont {Cho}, \citenamefont {Sangwan}, \citenamefont {Liu},
  \citenamefont {Lauhon}, \citenamefont {Marks},\ and\ \citenamefont
  {Hersam}}]{doi10.1021nl5032293}%
  \BibitemOpen
  \bibfield  {author} {\bibinfo {author} {\bibfnamefont {J.~D.}\ \bibnamefont
  {Wood}}, \bibinfo {author} {\bibfnamefont {S.~A.}\ \bibnamefont {Wells}},
  \bibinfo {author} {\bibfnamefont {D.}~\bibnamefont {Jariwala}}, \bibinfo
  {author} {\bibfnamefont {K.-S.}\ \bibnamefont {Chen}}, \bibinfo {author}
  {\bibfnamefont {E.}~\bibnamefont {Cho}}, \bibinfo {author} {\bibfnamefont
  {V.~K.}\ \bibnamefont {Sangwan}}, \bibinfo {author} {\bibfnamefont
  {X.}~\bibnamefont {Liu}}, \bibinfo {author} {\bibfnamefont {L.~J.}\
  \bibnamefont {Lauhon}}, \bibinfo {author} {\bibfnamefont {T.~J.}\
  \bibnamefont {Marks}}, \ and\ \bibinfo {author} {\bibfnamefont {M.~C.}\
  \bibnamefont {Hersam}},\ }\href {\doibase 10.1021/nl5032293} {\bibfield
  {journal} {\bibinfo  {journal} {Nano Letters}\ }\textbf {\bibinfo {volume}
  {14}},\ \bibinfo {pages} {6964} (\bibinfo {year} {2014})},\ \bibinfo {note}
  {pMID: 25380142},\ \Eprint
  {http://arxiv.org/abs/http://dx.doi.org/10.1021/nl5032293}
  {http://dx.doi.org/10.1021/nl5032293} \BibitemShut {NoStop}%
\bibitem [{\citenamefont {Kulish}\ \emph {et~al.}(2015)\citenamefont {Kulish},
  \citenamefont {Malyi}, \citenamefont {Persson},\ and\ \citenamefont
  {Wu}}]{C4CP03890H}%
  \BibitemOpen
  \bibfield  {author} {\bibinfo {author} {\bibfnamefont {V.~V.}\ \bibnamefont
  {Kulish}}, \bibinfo {author} {\bibfnamefont {O.~I.}\ \bibnamefont {Malyi}},
  \bibinfo {author} {\bibfnamefont {C.}~\bibnamefont {Persson}}, \ and\
  \bibinfo {author} {\bibfnamefont {P.}~\bibnamefont {Wu}},\ }\href {\doibase
  10.1039/C4CP03890H} {\bibfield  {journal} {\bibinfo  {journal} {Phys. Chem.
  Chem. Phys.}\ }\textbf {\bibinfo {volume} {17}},\ \bibinfo {pages} {992}
  (\bibinfo {year} {2015})}\BibitemShut {NoStop}%
\bibitem [{\citenamefont {Slack}(1965)}]{PhysRev.139.A507}%
  \BibitemOpen
  \bibfield  {author} {\bibinfo {author} {\bibfnamefont {G.~A.}\ \bibnamefont
  {Slack}},\ }\href {\doibase 10.1103/PhysRev.139.A507} {\bibfield  {journal}
  {\bibinfo  {journal} {Phys. Rev.}\ }\textbf {\bibinfo {volume} {139}},\
  \bibinfo {pages} {A507} (\bibinfo {year} {1965})}\BibitemShut {NoStop}%
\bibitem [{\citenamefont {Kresse}\ and\ \citenamefont
  {Furthm\"uller}(1996)}]{PhysRevB.54.11169}%
  \BibitemOpen
  \bibfield  {author} {\bibinfo {author} {\bibfnamefont {G.}~\bibnamefont
  {Kresse}}\ and\ \bibinfo {author} {\bibfnamefont {J.}~\bibnamefont
  {Furthm\"uller}},\ }\href {\doibase 10.1103/PhysRevB.54.11169} {\bibfield
  {journal} {\bibinfo  {journal} {Phys. Rev. B}\ }\textbf {\bibinfo {volume}
  {54}},\ \bibinfo {pages} {11169} (\bibinfo {year} {1996})}\BibitemShut
  {NoStop}%
\bibitem [{\citenamefont {Perdew}\ \emph {et~al.}(1996)\citenamefont {Perdew},
  \citenamefont {Burke},\ and\ \citenamefont
  {Ernzerhof}}]{PhysRevLett.77.3865}%
  \BibitemOpen
  \bibfield  {author} {\bibinfo {author} {\bibfnamefont {J.~P.}\ \bibnamefont
  {Perdew}}, \bibinfo {author} {\bibfnamefont {K.}~\bibnamefont {Burke}}, \
  and\ \bibinfo {author} {\bibfnamefont {M.}~\bibnamefont {Ernzerhof}},\ }\href
  {\doibase 10.1103/PhysRevLett.77.3865} {\bibfield  {journal} {\bibinfo
  {journal} {Phys. Rev. Lett.}\ }\textbf {\bibinfo {volume} {77}},\ \bibinfo
  {pages} {3865} (\bibinfo {year} {1996})}\BibitemShut {NoStop}%
\bibitem [{\citenamefont {Monkhorst}\ and\ \citenamefont
  {Pack}(1976)}]{PhysRevB.13.5188}%
  \BibitemOpen
  \bibfield  {author} {\bibinfo {author} {\bibfnamefont {H.~J.}\ \bibnamefont
  {Monkhorst}}\ and\ \bibinfo {author} {\bibfnamefont {J.~D.}\ \bibnamefont
  {Pack}},\ }\href {\doibase 10.1103/PhysRevB.13.5188} {\bibfield  {journal}
  {\bibinfo  {journal} {Phys. Rev. B}\ }\textbf {\bibinfo {volume} {13}},\
  \bibinfo {pages} {5188} (\bibinfo {year} {1976})}\BibitemShut {NoStop}%
\bibitem [{\citenamefont {Lindsay}\ \emph {et~al.}(2010)\citenamefont
  {Lindsay}, \citenamefont {Broido},\ and\ \citenamefont
  {Mingo}}]{PhysRevB.82.115427}%
  \BibitemOpen
  \bibfield  {author} {\bibinfo {author} {\bibfnamefont {L.}~\bibnamefont
  {Lindsay}}, \bibinfo {author} {\bibfnamefont {D.~A.}\ \bibnamefont {Broido}},
  \ and\ \bibinfo {author} {\bibfnamefont {N.}~\bibnamefont {Mingo}},\ }\href
  {\doibase 10.1103/PhysRevB.82.115427} {\bibfield  {journal} {\bibinfo
  {journal} {Phys. Rev. B}\ }\textbf {\bibinfo {volume} {82}},\ \bibinfo
  {pages} {115427} (\bibinfo {year} {2010})}\BibitemShut {NoStop}%
\bibitem [{\citenamefont {Castro~Neto}\ \emph {et~al.}(2009)\citenamefont
  {Castro~Neto}, \citenamefont {Guinea}, \citenamefont {Peres}, \citenamefont
  {Novoselov},\ and\ \citenamefont {Geim}}]{RevModPhys.81.109}%
  \BibitemOpen
  \bibfield  {author} {\bibinfo {author} {\bibfnamefont {A.~H.}\ \bibnamefont
  {Castro~Neto}}, \bibinfo {author} {\bibfnamefont {F.}~\bibnamefont {Guinea}},
  \bibinfo {author} {\bibfnamefont {N.~M.~R.}\ \bibnamefont {Peres}}, \bibinfo
  {author} {\bibfnamefont {K.~S.}\ \bibnamefont {Novoselov}}, \ and\ \bibinfo
  {author} {\bibfnamefont {A.~K.}\ \bibnamefont {Geim}},\ }\href {\doibase
  10.1103/RevModPhys.81.109} {\bibfield  {journal} {\bibinfo  {journal} {Rev.
  Mod. Phys.}\ }\textbf {\bibinfo {volume} {81}},\ \bibinfo {pages} {109}
  (\bibinfo {year} {2009})}\BibitemShut {NoStop}%
\bibitem [{\citenamefont {Zhang}\ \emph {et~al.}(2014)\citenamefont {Zhang},
  \citenamefont {Xie}, \citenamefont {Hu}, \citenamefont {Bao}, \citenamefont
  {Yue}, \citenamefont {Qin},\ and\ \citenamefont {Su}}]{PhysRevB.89.054310}%
  \BibitemOpen
  \bibfield  {author} {\bibinfo {author} {\bibfnamefont {X.}~\bibnamefont
  {Zhang}}, \bibinfo {author} {\bibfnamefont {H.}~\bibnamefont {Xie}}, \bibinfo
  {author} {\bibfnamefont {M.}~\bibnamefont {Hu}}, \bibinfo {author}
  {\bibfnamefont {H.}~\bibnamefont {Bao}}, \bibinfo {author} {\bibfnamefont
  {S.}~\bibnamefont {Yue}}, \bibinfo {author} {\bibfnamefont {G.}~\bibnamefont
  {Qin}}, \ and\ \bibinfo {author} {\bibfnamefont {G.}~\bibnamefont {Su}},\
  }\href {\doibase 10.1103/PhysRevB.89.054310} {\bibfield  {journal} {\bibinfo
  {journal} {Phys. Rev. B}\ }\textbf {\bibinfo {volume} {89}},\ \bibinfo
  {pages} {054310} (\bibinfo {year} {2014})}\BibitemShut {NoStop}%
\bibitem [{\citenamefont {Xie}\ \emph {et~al.}(2014)\citenamefont {Xie},
  \citenamefont {Hu},\ and\ \citenamefont {Bao}}]{apl1041310.10631.4870586}%
  \BibitemOpen
  \bibfield  {author} {\bibinfo {author} {\bibfnamefont {H.}~\bibnamefont
  {Xie}}, \bibinfo {author} {\bibfnamefont {M.}~\bibnamefont {Hu}}, \ and\
  \bibinfo {author} {\bibfnamefont {H.}~\bibnamefont {Bao}},\ }\href {\doibase
  http://dx.doi.org/10.1063/1.4870586} {\bibfield  {journal} {\bibinfo
  {journal} {Appl. Phys. Lett.}\ }\textbf {\bibinfo {volume} {104}},\ \bibinfo
  {eid} {131906} (\bibinfo {year} {2014})}\BibitemShut {NoStop}%
\bibitem [{\citenamefont {Fei}\ and\ \citenamefont
  {Yang}(2014)}]{doi:10.1021/nl500935z}%
  \BibitemOpen
  \bibfield  {author} {\bibinfo {author} {\bibfnamefont {R.}~\bibnamefont
  {Fei}}\ and\ \bibinfo {author} {\bibfnamefont {L.}~\bibnamefont {Yang}},\
  }\href {\doibase 10.1021/nl500935z} {\bibfield  {journal} {\bibinfo
  {journal} {Nano Lett.}\ }\textbf {\bibinfo {volume} {14}},\ \bibinfo {pages}
  {2884} (\bibinfo {year} {2014})}\BibitemShut {NoStop}%
\bibitem [{\citenamefont {Togo}\ \emph {et~al.}(2008)\citenamefont {Togo},
  \citenamefont {Oba},\ and\ \citenamefont {Tanaka}}]{phonopy}%
  \BibitemOpen
  \bibfield  {author} {\bibinfo {author} {\bibfnamefont {A.}~\bibnamefont
  {Togo}}, \bibinfo {author} {\bibfnamefont {F.}~\bibnamefont {Oba}}, \ and\
  \bibinfo {author} {\bibfnamefont {I.}~\bibnamefont {Tanaka}},\ }\href@noop {}
  {\bibfield  {journal} {\bibinfo  {journal} {Phys. Rev. B}\ }\textbf {\bibinfo
  {volume} {78}},\ \bibinfo {pages} {134106} (\bibinfo {year}
  {2008})}\BibitemShut {NoStop}%
\bibitem [{\citenamefont {Broido}\ \emph {et~al.}(2007)\citenamefont {Broido},
  \citenamefont {Malorny}, \citenamefont {Birner}, \citenamefont {Mingo},\ and\
  \citenamefont {Stewart}}]{ap1.2822891}%
  \BibitemOpen
  \bibfield  {author} {\bibinfo {author} {\bibfnamefont {D.~A.}\ \bibnamefont
  {Broido}}, \bibinfo {author} {\bibfnamefont {M.}~\bibnamefont {Malorny}},
  \bibinfo {author} {\bibfnamefont {G.}~\bibnamefont {Birner}}, \bibinfo
  {author} {\bibfnamefont {N.}~\bibnamefont {Mingo}}, \ and\ \bibinfo {author}
  {\bibfnamefont {D.~A.}\ \bibnamefont {Stewart}},\ }\href {\doibase
  http://dx.doi.org/10.1063/1.2822891} {\bibfield  {journal} {\bibinfo
  {journal} {Appl. Phys. Lett.}\ }\textbf {\bibinfo {volume} {91}},\ \bibinfo
  {eid} {231922} (\bibinfo {year} {2007})}\BibitemShut {NoStop}%
\bibitem [{\citenamefont {Li}\ \emph {et~al.}(2012)\citenamefont {Li},
  \citenamefont {Lindsay}, \citenamefont {Broido}, \citenamefont {Stewart},\
  and\ \citenamefont {Mingo}}]{PhysRevB.86.174307}%
  \BibitemOpen
  \bibfield  {author} {\bibinfo {author} {\bibfnamefont {W.}~\bibnamefont
  {Li}}, \bibinfo {author} {\bibfnamefont {L.}~\bibnamefont {Lindsay}},
  \bibinfo {author} {\bibfnamefont {D.~A.}\ \bibnamefont {Broido}}, \bibinfo
  {author} {\bibfnamefont {D.~A.}\ \bibnamefont {Stewart}}, \ and\ \bibinfo
  {author} {\bibfnamefont {N.}~\bibnamefont {Mingo}},\ }\href {\doibase
  10.1103/PhysRevB.86.174307} {\bibfield  {journal} {\bibinfo  {journal} {Phys.
  Rev. B}\ }\textbf {\bibinfo {volume} {86}},\ \bibinfo {pages} {174307}
  (\bibinfo {year} {2012})}\BibitemShut {NoStop}%
\bibitem [{\citenamefont {Lindsay}\ \emph {et~al.}(2014)\citenamefont
  {Lindsay}, \citenamefont {Li}, \citenamefont {Carrete}, \citenamefont
  {Mingo}, \citenamefont {Broido},\ and\ \citenamefont
  {Reinecke}}]{PhysRevB.89.155426}%
  \BibitemOpen
  \bibfield  {author} {\bibinfo {author} {\bibfnamefont {L.}~\bibnamefont
  {Lindsay}}, \bibinfo {author} {\bibfnamefont {W.}~\bibnamefont {Li}},
  \bibinfo {author} {\bibfnamefont {J.}~\bibnamefont {Carrete}}, \bibinfo
  {author} {\bibfnamefont {N.}~\bibnamefont {Mingo}}, \bibinfo {author}
  {\bibfnamefont {D.~A.}\ \bibnamefont {Broido}}, \ and\ \bibinfo {author}
  {\bibfnamefont {T.~L.}\ \bibnamefont {Reinecke}},\ }\href {\doibase
  10.1103/PhysRevB.89.155426} {\bibfield  {journal} {\bibinfo  {journal} {Phys.
  Rev. B}\ }\textbf {\bibinfo {volume} {89}},\ \bibinfo {pages} {155426}
  (\bibinfo {year} {2014})}\BibitemShut {NoStop}%
\bibitem [{\citenamefont {Li}\ \emph {et~al.}(2014{\natexlab{b}})\citenamefont
  {Li}, \citenamefont {Carrete}, \citenamefont {Katcho},\ and\ \citenamefont
  {Mingo}}]{Li20141747}%
  \BibitemOpen
  \bibfield  {author} {\bibinfo {author} {\bibfnamefont {W.}~\bibnamefont
  {Li}}, \bibinfo {author} {\bibfnamefont {J.}~\bibnamefont {Carrete}},
  \bibinfo {author} {\bibfnamefont {N.~A.}\ \bibnamefont {Katcho}}, \ and\
  \bibinfo {author} {\bibfnamefont {N.}~\bibnamefont {Mingo}},\ }\href
  {\doibase http://dx.doi.org/10.1016/j.cpc.2014.02.015} {\bibfield  {journal}
  {\bibinfo  {journal} {Comput. Phys. Commun.}\ }\textbf {\bibinfo {volume}
  {185}},\ \bibinfo {pages} {1747 } (\bibinfo {year}
  {2014}{\natexlab{b}})}\BibitemShut {NoStop}%
\bibitem [{\citenamefont {Carrete}\ \emph {et~al.}(2014)\citenamefont
  {Carrete}, \citenamefont {Mingo},\ and\ \citenamefont
  {Curtarolo}}]{apl.10510101907.1.4895770}%
  \BibitemOpen
  \bibfield  {author} {\bibinfo {author} {\bibfnamefont {J.}~\bibnamefont
  {Carrete}}, \bibinfo {author} {\bibfnamefont {N.}~\bibnamefont {Mingo}}, \
  and\ \bibinfo {author} {\bibfnamefont {S.}~\bibnamefont {Curtarolo}},\ }\href
  {\doibase http://dx.doi.org/10.1063/1.4895770} {\bibfield  {journal}
  {\bibinfo  {journal} {Appl. Phys. Lett.}\ }\textbf {\bibinfo {volume}
  {105}},\ \bibinfo {eid} {101907} (\bibinfo {year} {2014})}\BibitemShut
  {NoStop}%
\bibitem [{\citenamefont {Parker}\ and\ \citenamefont
  {Singh}(2010)}]{PhysRevB.82.035204}%
  \BibitemOpen
  \bibfield  {author} {\bibinfo {author} {\bibfnamefont {D.}~\bibnamefont
  {Parker}}\ and\ \bibinfo {author} {\bibfnamefont {D.~J.}\ \bibnamefont
  {Singh}},\ }\href {\doibase 10.1103/PhysRevB.82.035204} {\bibfield  {journal}
  {\bibinfo  {journal} {Phys. Rev. B}\ }\textbf {\bibinfo {volume} {82}},\
  \bibinfo {pages} {035204} (\bibinfo {year} {2010})}\BibitemShut {NoStop}%
\bibitem [{\citenamefont {Pulikkotil}\ \emph {et~al.}(2012)\citenamefont
  {Pulikkotil}, \citenamefont {Singh}, \citenamefont {Auluck}, \citenamefont
  {Saravanan}, \citenamefont {Misra}, \citenamefont {Dhar},\ and\ \citenamefont
  {Budhani}}]{PhysRevB.86.155204}%
  \BibitemOpen
  \bibfield  {author} {\bibinfo {author} {\bibfnamefont {J.~J.}\ \bibnamefont
  {Pulikkotil}}, \bibinfo {author} {\bibfnamefont {D.~J.}\ \bibnamefont
  {Singh}}, \bibinfo {author} {\bibfnamefont {S.}~\bibnamefont {Auluck}},
  \bibinfo {author} {\bibfnamefont {M.}~\bibnamefont {Saravanan}}, \bibinfo
  {author} {\bibfnamefont {D.~K.}\ \bibnamefont {Misra}}, \bibinfo {author}
  {\bibfnamefont {A.}~\bibnamefont {Dhar}}, \ and\ \bibinfo {author}
  {\bibfnamefont {R.~C.}\ \bibnamefont {Budhani}},\ }\href {\doibase
  10.1103/PhysRevB.86.155204} {\bibfield  {journal} {\bibinfo  {journal} {Phys.
  Rev. B}\ }\textbf {\bibinfo {volume} {86}},\ \bibinfo {pages} {155204}
  (\bibinfo {year} {2012})}\BibitemShut {NoStop}%
\bibitem [{\citenamefont {Tritt}(2004)}]{bookTCPOSALT}%
  \BibitemOpen
  \bibfield  {author} {\bibinfo {author} {\bibfnamefont {T.}~\bibnamefont
  {Tritt}},\ }in\ \href@noop {} {\emph {\bibinfo {booktitle} {Thermal
  conductivity: theory, properties, and applications}}}\ (\bibinfo  {publisher}
  {Kluwer Academic / Plenum Publishers},\ \bibinfo {year} {2004})\ pp.\
  \bibinfo {pages} {114--115}\BibitemShut {NoStop}%
\bibitem [{\citenamefont {Hu}\ \emph {et~al.}(2013)\citenamefont {Hu},
  \citenamefont {Zhang},\ and\ \citenamefont
  {Poulikakos}}]{PhysRevB.87.195417}%
  \BibitemOpen
  \bibfield  {author} {\bibinfo {author} {\bibfnamefont {M.}~\bibnamefont
  {Hu}}, \bibinfo {author} {\bibfnamefont {X.}~\bibnamefont {Zhang}}, \ and\
  \bibinfo {author} {\bibfnamefont {D.}~\bibnamefont {Poulikakos}},\ }\href
  {\doibase 10.1103/PhysRevB.87.195417} {\bibfield  {journal} {\bibinfo
  {journal} {Phys. Rev. B}\ }\textbf {\bibinfo {volume} {87}},\ \bibinfo
  {pages} {195417} (\bibinfo {year} {2013})}\BibitemShut {NoStop}%
\bibitem [{\citenamefont {Gu}\ and\ \citenamefont
  {Yang}(2014)}]{arXiv:1404.2874}%
  \BibitemOpen
  \bibfield  {author} {\bibinfo {author} {\bibfnamefont {X.}~\bibnamefont
  {Gu}}\ and\ \bibinfo {author} {\bibfnamefont {R.}~\bibnamefont {Yang}},\
  }\href@noop {} {\bibfield  {journal} {\bibinfo  {journal} {arXiv:1404.2874}\
  } (\bibinfo {year} {2014})}\BibitemShut {NoStop}%
\end{thebibliography}%

\end{document}